\documentclass[12pt,preprint]{aastex}
\begin{document}

\markboth{S.Tanuma \& K.Shibata  2004, Submitted to Astrophysical Journal}
{Internal Shocks in the Magnetic Reconnection Jet in Solar Flares}

\title{
Internal Shocks in the Magnetic Reconnection Jet in Solar Flares:
Multiple Fast Shocks Created by the Secondary Tearing Instability}

\author{\center 
S.\ Tanuma\altaffilmark{1}
\& K.\ Shibata\altaffilmark{1}}

\altaffiltext{1}{Kwasan Observatory,
Kyoto University, Yamashina, Kyoto, 607-8471, Japan:
tanuma@kwasan.kyoto-u.ac.jp, shibata@kwasan.kyoto-u.ac.jp}

\begin{abstract}
Space solar missions such as {\it Yohkoh} and {\it RHESSI} observe 
the hard X- and gamma-ray emission from energetic electrons
in impulsive solar flares.
Their energization mechanism, however, is unknown.
In this paper,
we suggest that the internal shocks are created in the reconnection jet
and that they are possible sites of particle acceleration.
We examine how magnetic reconnection creates the multiple shocks
by performing two-dimensional resistive magnetohydrodynamic simulations.
In this paper, we use a very small grid to resolve the diffusion region.
As a result, we find that 
the current sheet becomes thin due to the tearing instability,
and it collapses to a Sweet-Parker sheet.
The thin sheet becomes unstable to the secondary tearing instability.
Fast reconnection starts by the onset of anomalous resistivity
immediately after the secondary tearing instability.
During the bursty, time-dependent magnetic reconnection,
the secondary tearing instability continues in the diffusion region
where the anomalous resistivity is enhanced.
As a result, many weak shocks are created in the reconnection jet.
This situation produces turbulent reconnection.
We suggest that
multiple fast shocks are created in the jet
and that the energetic electrons can be accelerated by these shocks.
\\
\end{abstract}


\section{INTRODUCTION}

In impulsive flares, 
strong hard X-ray emission from energetic electrons
has been observed by {\it Yohkoh} HXT (Masuda et al.\ 1994; 
see also a review by Miller et al.\ 1997).
The origin of energetic electrons is, however, unknown.
The fast-mode shock is known to be a possible site
for particle acceleration, 
for example in interplanetary space \citep{ter89}
and supernova remnants \citep{bla80,koy95}.
To solve the particle acceleration problem in solar flares,
\citet{tsu98} suggested a diffusive shock model at the loop top. 
This model, however, has not yet been fully examined.

Solar flares occur as a result of magnetic reconnection.
The SXT and HXT on the {\it Yohkoh} satellite discovered 
various pieces of evidence of magnetic reconnection
(Tsuneta et al.\ 1992; Masuda et al.\ 1994; see review by Shibata 1999).
The ``fast'' reconnection process suggested by \citet{pet64}
can be supported by a localized resistivity
(e.g, anomalous resistivity;
Ugai 1986; Tanuma et al.\ 2003; 
see also Shibata, Nozawa, \& Matsumoto 1992; Miyagoshi \& Yokoyama 2003).
Recently, \citet{tan99,tan01} investigated
the reconnection triggered by a shock wave from a point explosion.
They found that Petschek reconnection occurs
immediately after the ``secondary tearing instability'' 
(Furth, Killeen, \& Rosenbluth 1963;
see also Magara \& Shibata 1999; Shibata \& Tanuma 2001),
and that fast shocks are created in the reconnection jet.
In this paper, we suggest that 
energetic electrons could be accelerated
if the internal fast shocks could be created in solar flares
(figure \ref{illust}).
To examine this possibility,
we study how the internal shocks are created in the reconnection jet
by performing 2D  simulations with high spatial resolution.
This should remove the effect of numerical noise 
and resolve the diffusion region.

These simulations lead us to propose that
multiple fast shocks are created in solar flares
and they are possible sites of particle acceleration.
Our simulation method is described in the next section.
In section 3, we present the simulation results.
Finally, we discuss the results.

\section{Model of Numerical Simulations}

We solve the nonlinear, time-dependent, resistive, 
compressible MHD equations.
A rectangular computation box with 2D Cartesian coordinates 
in the $x$-$z$ plane is assumed,
where the $x$ and $z$ axes are in the horizontal and vertical directions,
respectively.
We assume an ideal gas, i.e., $p_g=(\gamma-1)e$,
where $\gamma$ is the specific heat ratio (=5/3).
The velocity and magnetic field are
$\mbox{\boldmath$v$}=(v_x, v_z)$ and $\mbox{\boldmath$B$}=(B_x, B_z)$.
Gravity is neglected.

In the initial equilibrium conditions, 
a Harris current sheet is assumed, i.e.,
$\mbox{\boldmath$B$}(x,z)=B_0\tanh(z/l^{\rm init})\mbox{\boldmath$x$}$,
$p_g(x,z)=p_{g0}+(B_0^2/8\pi)[1-\tanh^2(z/l^{\rm init})]$,
$\rho(x,z)=(\gamma p_g/T)
=\rho_0+(\gamma/T_0)(B_0^2/8\pi)[1-\tanh^2(z/l^{\rm init})]$,
where $B_0$, $p_{g0}$, and $\rho_0$, are dimensionless variables, 
and $\mbox{\boldmath$x$}=(1,0)$.
We assume in the initial conditions
that the current-sheet half-thickness $l^{\rm init}=1$.
The ratio of gas to magnetic pressure is
$\beta=8\pi p_{g0}/B_0^2=0.2$ ($|z|\gg l^{\rm init}$).
The initial gas pressure is $p_{g0}$ outside the current sheet,
and $p_{g0}+B_0^2/8\pi=p_{g0}(1+1/\beta)$ inside the current sheet.
The total pressure is uniform.
We assume $p_{g0}=1/\gamma=0.6$ and $\rho_0=1$.
The sound velocity and temperature are
$C_s\equiv (\gamma p_g/\rho)^{1/2}=1$ (uniform) 
and $T=T_0=1$ (uniform).
The initial Alfv\'en velocity is
$v_{\rm A}^{\rm init}=B_0/(4\pi\rho_0)^{1/2}\simeq 2.45$
($|z|\gg l^{\rm init}$).
We assume an anomalous resistivity model as follows:
$\eta=\eta_0$ for $v_d\leq v_c$, and
$\eta=\eta_0+\alpha (v_d/v_c-1)^2$ for $v_d>v_c$
\citep{kli98,uga86,tan99,tan01},
where $v_d$($\equiv J/\rho$), $\rho$, $J$, and $v_c$
are the relative ion-electron drift velocity,
mass density, current density,
and threshold above which the anomalous resistivity sets in
(Anomalous resistivity originates from a lower hybrid instability,
ion-acoustic instability, or ion-cyclotron instability;
see, e.g., Treumann 2001).
We also assume that the resistivity does not exceed $\eta_{\rm max}=1$.
In this paper, we assume a ``background resistivity'' $\eta_0=0.005$,
which is sufficiently larger than the ``numerical resistivity'' 
because of the grid size \citep{uga99,tan01}.
The other parameters are $\alpha=10.0$ and $v_c=20.0$.
We neglect the thermal radiation and heat conduction.
They do not affect the basic properties of magnetic reconnection
\citep{yok97}.

We normalize the velocity, length, and time
by the sound velocity ($C_s$), initial current-sheet thickness ($H$),
and $H/C_s$, respectively.
The units of normalization are
$C_s\sim 150$ km s$^{-1}$,
$H\sim 3000$ km,
and $\tau\equiv H/C_s\sim 20$ s.
The units of temperature, density, gas pressure, 
and magnetic field strength are
$T_0\sim 2\times 10^6$ K,
$n_0\sim 10^9$ cm$^{-3}$,
$p_{g0}\sim 10^{-1}$ erg cm$^{-3}$,
and $B_0\sim 2$ G, respectively
($B_0$ is $\sim 10$ G if we assume $n_0\sim 2.5\times 10^{10}$ cm$^{-3}$).
The grid number is $(N_x, N_z)=(13000, 1300)$,
and the grid size is uniform $(\triangle x, \triangle z)=(0.013, 0.013)$.
We assume that
the top ($z=+8.45$) and bottom ($z=-8.45$) surfaces are free boundaries, 
and that the right ($x=+84.5$) and left ($x=-84.5$) ones are periodic.
The simulation box size is $(L_x, L_z)=(169.0, 16.9)$.
The magnetic Reynolds number is
${\rm Re}_m^{\rm init}\equiv v_{\rm A}^{\rm init}L_x/\eta_0
\sim 84500$.
We use a 2-step modified Lax-Wendroff method.
The resistivity is initially enhanced for a short time 
in the central region of the current sheet 
\citep[see also][]{uga86}.

\section{Results}

Figure \ref{shocks} shows the time variation of the spatial distribution
of the gas pressure in the reconnection region.
The current sheet is unstable because of the tearing instability 
(Furth et al. 1963).
The magnetic dissipation time, Alfv\'en time,
and tearing instability time scale are
$\tau_{\rm dis}^{\rm init}=l^{\rm init}{}^2/\eta_0
\sim 200$,
$\tau_{\rm A}^{\rm init}=l^{\rm init}/v_{\rm A}^{\rm init}
\sim 0.4$, and
$\tau_t^{\rm init}=(\tau_{\rm dis}^{\rm init}\tau_{\rm A}^{\rm init})^{1/2}
\sim 9$,
respectively,
where $l^{\rm init}(=1)$ is the half thickness of the initial current sheet.
The tearing instability is initiated in the current sheet
by the initial perturbation.
The current sheet becomes gradually thinner
in the nonlinear phase of the tearing instability ($t\sim 7-18$).
Its length is
comparable to the most unstable wavelength of the tearing instability,
i.e.,
$\lambda_t\sim 5.6{\rm Re}_{m,z}^{\rm init}{}^{1/4}l^{\rm init}$
\citep{mag99,tan01},
where
${\rm Re}_{m,z}^{\rm init}
\equiv v_{\rm A}^{\rm init}l^{\rm init}/\eta_0
\sim 500$.
$\lambda_t$ is $\sim 25$ in this simulation.
The current-sheet thickness in this phase is 
$l_t\sim (\lambda_t/2){\rm Re}_{m,t}^{-1/2}$,
i.e., the current sheet corresponds to 
that of a  Sweet(1958)-Parker(1957) sheet.
The magnetic Reynolds number is
${\rm Re}_{m,t}=(\lambda_t/2)v_{\rm A}^{\rm init}/\eta_0$ in this phase.
Furthermore, Re$_{m,t}\sim 6000$ and $l_t\sim 0.15$.
These values explain our results well.

The sheet becomes very long 
so that it becomes unstable to the tearing instability again at $t\sim 17-78$
\citep[referred to as the ``secondary tearing instability'' in][]{tan99,tan01}.
The secondary tearing instability starts
just after the Sweet-Parker sheet is created on a time scale of
$\tau_{2t}=(l_t{}^3/\eta_0v_{\rm A}^{\rm init})^{1/2}
\sim 0.52$.
The drift velocity ($v_d$) reaches a threshold ($v_c$) at $t\sim 18$
so that anomalous resistivity sets in.
$v_d$, however, remains a little larger than $v_c$ when $18<t<20$.
Figure \ref{tearing}a shows that the magnetic islands are actually created 
at the diffusion region
(But plasmoids are difficult to represent clearly in figure \ref{shocks}a).
The gaps between them ($\sim 0.3-0.5$) are not consistent with 
the analytic solution of the secondary tearing instability in this model,
$\lambda_{2t}\sim 4.9{{\rm Re}_{m,z}^{2t}}^{1/4}l_t$,
where ${\rm Re}_{m,z}^{2t}=v_{\rm A}^{\rm init}l_t/\eta_0$,
$\lambda_{2t}\sim 2.2$, and Re$_{m,z}^{2t}\sim 73.5$, respectively.
The current-sheet thickness becomes
$l_{2t}={{\rm Re}_m^{2t}}^{-1/2}(\lambda_{2t}/2) 
\sim 0.047$, 
where the magnetic Reynolds number is
Re$_m^{2t}=(\lambda_{2t}/2)v_{\rm A}^{\rm init}/\eta_0
=539$.
This is because 
the analytic solution is in the large ${\rm Re}_{m}$ approximation regime.

The drift velocity increases a great deal above the threshold at $t\sim 20$
before the current sheet completes its collapse to a thickness of $l_{2t}$.
Because of the excitation of anomalous resistivity,
non-steady Petschek(1964)-like (fast) reconnection starts,
and is accompanied by slow shocks.
During fast reconnection ($t>20$),
the secondary tearing instability occurs in the dissipation region
where the anomalous resistivity is enhanced.
This diffusion region is well resolved numerically.
The magnetic Reynolds number is now
${\rm Re}_{m,z}^{2t*}=v_{\rm A}^{\rm init}l_t/\eta_{\rm max}\sim 0.38$.
The wavelength and time scale of the secondary tearing instability are 
$\lambda_{2t}^*\sim 4.9{{\rm Re}_{m,z}^{2t*}}^{1/4}l_t\sim 0.58$
and
$\tau_{2t}^*=(l_t{}^3/\eta_{\rm max}v_{\rm A}^{\rm init})^{1/2}
\sim 0.0014$, respectively,
provided we assume that the secondary tearing instability starts 
when $l\sim l_t$.
Figure \ref{tearing}b also shows that the plasmoids are created
at this time.
Their wavelengths are consistent with the above theoretical value.
This result can be explained by extension of the results of \citet{ste83},
although the analytic solution is 
in the large ${\rm Re}_{m}$ approximation regime.
Due to the bursty, time-dependent reconnection,
many weak shocks are created in the reconnection jet.
They are almost stationary, 
when they are created outside the diffusion region,
although they move inside and near the diffusion region.
Figure \ref{profile} displays the profiles of some variables in $z=0.058$, 
when $t=35.0$.
The profiles show that 12 weak shocks are created among $0<x=6.5$.
Hence, the gaps between them ($\sim 6.5/12\sim 0.54$) 
are quite consistent with $\lambda_{2t}^*$
(Note that the local Alfv\'en velocity is smaller than $v_{\rm A}^{\rm init}$
during this phase).
The reconnection jet can become supersonic
and the pressure jumps are almost standing so that they could be shocks.
The reconnection rate and inflow velocity toward the diffusion region 
oscillate.
As a result, the reconnection becomes ``turbulent''
\citep{mat85, mat86, laz99, fan04} during the late phase.
In this phase, the Kelvin-Helmholtz instability may also occurs
(see discussion).

We examine the dependence of the results on the plasma $\beta$ 
and resistivity model ($\eta_0$, $v_c$, and $\alpha$). 
These parameters do not affect our principal results;
fast shocks are created in all simulations.
For example, the physical values are 
${\rm Re}_{m,z}^{\rm init}$
$\sim 500(l^{\rm init}/1.0)$
$(\eta_0/0.005)^{-1}$
$(\beta/0.2)^{-1/2}$,
$\lambda_t$
$\sim 25(l^{\rm init}/1.0)^{5/4}$
$(\eta_0/0.005)^{-1/4}$
$(\beta/0.2)^{-1/8}$,
and 
$l_t$
$\sim 0.15(l^{\rm init}/1.0)^{5/8}$
$(\eta_0/0.005)^{3/8}$
$(\beta/0.2)^{3/16}$, respectively.
Furthermore,
$\lambda_{2t}^*$
$\sim 0.58(l^{\rm init}/1.0)^{25/32}$
$(\eta_0/0.005)^{15/32}$
$(\eta_{\rm max}/1.0)^{-1/4}$
$(\beta/0.2)^{7/64}$,
and
$\tau_{2t}^*$
$\sim 0.0014(l^{\rm init}/1.0)^{15/16}$
$(\eta_0/0.005)^{9/16}$
$(\eta_{\rm max}/1.0)^{-1/2}$
$(\beta/0.2)^{17/32}$,
if anomalous resistivity is assumed to be excited to 
$\eta\sim\eta_{\rm max}$
when the current-sheet thickness is $l\sim l_t$.
A detailed investigation of the parameter-dependence will be published
elsewhere.

\section{Discussion}

We suggest that bursty, time-dependent reconnection creates 
many weak shocks (plasmoids) in the reconnection jet,
which could become multiple fast shocks in actual flares.
Reconnection produces an outflow 
whose velocity is comparable with the local sound speed.
Thus, it can create fast shocks 
which would provide sites for Fermi acceleration.
From our numerical simulations,
the number of energetic electrons can be
determined by the flux into the internal shocks,
and is calculated to be
$dN/dt\sim 10^{35}$ s$^{-1}$ \citep[see also][]{tsu98}.
The reconnection site evolves in a self-similar manner \citep{nit01},
so the evidence of bursty reconnection and internal shocks
shown in this paper may be observed by 
the {\it Solar-B} satellite which will be launched in 2006.
If the energetic electrons are accelerated in the fast shocks,
the number of energetic electrons would increase with distance from the
diffusion region.

Our simulations show 
that the reconnection jet (``downflow'') produces turbulence.
The turbulent field at the loop top helps
to accelerate and confine the energetic electrons \citep{jak98}.
As a second important application,
we suggest that the non-steady ejection of plasmoids 
can create the internal shocks.
They actually appear in some of our simulations with different parameters,
which we will examine in a future paper.
This could explain the oscillations \citep{inn03}
and non-steady plasmoid ejections \citep{mck99,asa04} 
in the downflows.
In fact, some results with different parameters show 
the oscillation and multiple oblique shocks in the jet,
which could be due to the Kelvin-Helmholtz instability 
\citep{bis98,tan00,arz01,tan02a,tan03b,tan04}.
Furthermore, 
spatial patterns that are similar to the multiple oblique shocks
appear in galactic jets \citep[e.g.,][]{har91}.

The non-steady ejection and bow shock propagation
can also play an important role in 
slowly drifting structures \citep{kar03}, 
and narrowband dm-spikes \citep{bar01} in the upward jet.
The time scale for the tearing instability in the diffusion region
is $\sim 30$ $(l/10^5\ {\rm cm})^{3/2}$ 
$(v_{\rm A}/10^8\ {\rm cm\ s^{-1}})^{-1/2}$
$(T/10^6\ {\rm K})^{3/4}$ s
if we assume Spitzer conductivity 
and that the diffusion region thickness is $l$.
This explains the time scales associated with the above phenomena 
\citep{kar03}.
Our results can also be applied to particle acceleration in 
the Galactic plane \citep{tan03a} etc.

\acknowledgments

The authors thank the anonymous referee for useful suggestions,
which improved this work very much.
One of the authors (S.T.) thanks 
M. Fujimoto,
M. Hoshino, 
B. Kliem,
K. Kondoh, 
Tohru Shimizu, 
A. Takeuchi,
T. Terasawa, 
S. Tsuneta, 
M. Ugai, \& 
T. Yokoyama
for various fruitful discussions and comments.
The authors also thank D. Brooks for brushing up this paper.
Part of the numerical computations were carried out 
on the VPP5000 at the ADAC of the NAOJ
(PI: S.\ Tanuma; project ID: mst21a, yst15a, rst03a).
The other part was carried out 
on the VPP5000 at the ITC, Nagoya Univ.,
by the joint research program of the STEL, Nagoya University (PI: S.\ Tanuma).
This work was partially supported by 
ACT-JST Cooperation (PI: R.\ Matsumoto),
by the JSPS Japan-US Cooperation Science Program 
(PI: K.\ Shibata \& K.\ I.\ Nishikawa), 
by the JSPS Japan-UK Cooperation Science Program
(PI: K.\ Shibata \& N.\ O.\ Weiss),
and the Grant-in-Aid for 
the 21st Century COE "Center for Diversity and Universality in Physics" 
from the MEXT
of Japan. 
S.T. is supported by a Grant-in-Aid for JSPS Fellows.

\clearpage
\begin{figure}[tbh]
\caption{
Schematic illustration of the internal shocks in the reconnection jet 
in solar flares.
\label{illust}
}
\end{figure}

\clearpage
\begin{figure}[tbh]
\caption{
The gas pressure.
The drift velocity reaches the threshold at $t\sim 18$.
Petschek-like reconnection starts at $t\sim 20$. 
Bursty, time-dependent fast reconnection creates many weak shocks 
in the reconnection jet.
\label{shocks}
}
\end{figure}

\clearpage
\begin{figure}[tbh]
\caption{
(a)
$B_z$ and $\mbox{\boldmath$B$}$ vectors in diffusion regions.
The white circles indicate the large $B_z$ regions.
It shows that small magnetic islands are created,
although their gas is smaller than 
the theoretical wavelength of the secondary tearing instability
($\lambda_{2t}\sim 2.2$).
(b)
The gas pressure and $\mbox{\boldmath$v$}$ vectors at the diffusion region.
The gaps between the shocks are quite consistent with the theoretical value
($\lambda^*_{2t}\sim 0.58$; shown by the arrow).
\label{tearing}
}
\end{figure}

\clearpage
\begin{figure}[tbh]
\caption{\small{
The profiles of some variables in $z=0.032$ at $t=35.0$
between '$\alpha$' and '$\beta$' plotted in fig \ref{shocks}.
Gray boxes show the pressure jumps created in the reconnection jet.
They are almost standing shocks.
The Rankine-Hugoniot relation is satisfied, for example, 
by more than 91 \% at the jump named 'B'.
In the 3rd plot, 
the local sound velocity ($C_s$) is also displayed by dashed-dotted lines.
The velocity of jet is comparable with $|C_s|$
so that the jet can be supersonic and can be weak shocks.
The pressure jumps are resolved by more than 10 grids.
The bottom plot is same with 3rd one but in $z=0.025$ at $t=60$.
label{profile}
}}
\end{figure}

\end{document}